\documentclass[12pt,preprint]{emulateapj}
\usepackage{graphicx}
\usepackage{epsfig}

\def\kms{km~s$^{-1}$}

\def\farcs{\hbox{$.\!\!^{\prime\prime}$}}

\slugcomment{Published in the Astrophysical Journal, 677, 306.}

\shorttitle{The Nature of SNR 4449-1}
\shortauthors{Milisavljevic \& Fesen}

\begin{document}

\title{The Nature of the Ultraluminous Oxygen-Rich \\
Supernova Remnant in NGC 4449}  

\author{Dan Milisavljevic \& Robert A.\ Fesen}

\affil{6127 Wilder Lab, Department of Physics \& Astronomy, Dartmouth
  College, Hanover, NH 03755 \\ danmil@dartmouth.edu, fesen@snr.darmouth.edu} 

\begin{abstract}

Optical images and spectra both ground-based and taken by the {\it
Hubble Space Telescope} ({\it HST}) of the young, luminous O-rich
supernova remnant in the irregular galaxy NGC 4449 are
presented. {\it HST} images of the remnant and its local region were
obtained with the ACS/WFC using filters F435W, F555W, F814W ($B$, $V$, and
$I$, respectively), F502N ([\ion{O}{3}]), F658N (H$\alpha$ +
[\ion{N}{2}]), F660N ([\ion{N}{2}]) and F550M (line-free
continuum). These images show an unresolved remnant (FWHM $<$ 0.05
arcsec) located within a rich cluster of OB stars which itself is
enclosed by a nearly complete interstellar shell seen best in
H$\alpha$ + [\ion{N}{2}] emission approximately 8$''$ $\times$ 6$''$
(150 $\times$ 110 pc) in size.  The remnant and its associated OB
cluster are isolated from two large nearby \ion{H}{2} regions.  The ACS
[\ion{O}{3}] image shows the remnant may be partially surrounded by a
clumpy ring of emission approximately 1$''$ ($\sim 20$ pc) in
diameter.  Recent ground-based spectra of the remnant reveal (1) the
emergence of broad, blueshifted emission lines of [\ion{S}{2}]
$\lambda\lambda$6716,6731, [\ion{Ar}{3}] $\lambda$7136, and
[\ion{Ca}{2}] $\lambda\lambda$7291,7324 which were not observed in
spectra taken in $1978-80$, (2) faint emission at 6540 --
6605 \AA \ centered about H$\alpha$ and [\ion{N}{2}]
$\lambda\lambda$6548,6583 with an expansion velocity of 500 $\pm$100
km s$^{-1}$, and (3) excess emission around 4600 -- 4700 \AA \
suggestive of a Wolf-Rayet population in the remnant's star cluster.
We use these new data to re-interpret the origin of the remnant's
prolonged and bright luminosity and propose the remnant is
strongly interacting with dense, circumstellar wind loss material from
a $\ga$ 20 M$_{\odot}$ progenitor star.  

\end{abstract}


\section{Introduction}

NGC 4449 is a Magellanic irregular type galaxy located at an estimated
distance of 3.82 $\pm$0.18 Mpc \citep{Annibali07}. Situated in the
northern outskirts of the galaxy, near two prominent H~II regions,
lies a young, oxygen-rich supernova remnant (SNR), hereafter SNR
4449-1.  With an observed radio flux of 4 mJy at 4.8 GHz (2002 epoch;
\citealt{Lacey07}) and an X-ray luminosity of  $2.4 \times 10^{38}$
erg s$^{-1}$ \citep{Patnaude03}, the remnant ranks among the most
luminous remnants known and is nearly an order of magnitude brighter
than the most luminous SNR in our galaxy, Cas A. 
 
The SNR was first discovered in the radio by \citet{Seaquist78} as a
bright, unresolved non-thermal radio source ($\sim$ 10 mJy at 2.7 GHz)
approximately 1$'$ north of the nucleus of the galaxy at a location
nearly coincident with an \ion{H}{2} region cataloged by
\citet{Sabbadin79}. Subsequent optical spectrophotometry showed two
different types of line emissions: (1) narrow \ion{H}{2} region-like
lines, and (2) broad lines of forbidden oxygen attributable to a young
SNR like Cas A \citep{Balick78}.  

Further studies in the optical, UV, and X-rays confirmed the SNR
nature of the object and highlighted its unusual luminousity and
relatively young age (\citealt{Kirshner80,Blair83,Blair84}).  
Optical spectra showed no broad hydrogen emission, yet very broad
emission of [\ion{O}{1}] $\lambda\lambda$6300,6364, [\ion{O}{2}]
$\lambda\lambda$7320,7330, and [\ion{O}{3}] $\lambda\lambda$4959,5007
suggestive of O-rich SN ejecta with an expansion velocity of
approximately 3500 km s$^{-1}$ and emitting an observed flux
greater than 3 $\times$ 10$^{-13}$ erg s$^{-1}$
cm$^{-2}$. \citet{Blair83} concluded that the remnant was
approximately 100 -- 200 years old and used several lines of reasoning
to infer a massive progenitor star $\approx$ 25 M$_{\odot}$.  Recent
work in the X-ray has suggested an age as old as $\sim$ 380 yr
\citep{Summers03}.   

The remnant appeared unresolved in ground-based optical images,
and VLBI measurements by \citet{deBruyn83} gave an upper limit of the
remnant's angular diameter of $\leq$ 0$\farcs$07 (1.3 pc).  A
subsequent analysis of optical spectra taken in 1993 using the Faint
Object Spectrograph (FOS) and images taken in 1996 with the Faint
Object Camera (FOC) onboard the {\it Hubble Space Telescope} ({\it
HST}) led \citet{Blair98} to report expansion velocities of the
remnant as high as 6000 km s$^{-1}$ and place an upper limit of
0$\farcs028$ on its diameter. This expansion velocity and size
suggested an age around 100 yr.  

While SNR 4449-1 is currently quite luminous, the significant decline
in the remnant's X-ray and radio flux over the last three decades
implies that it was even brighter in the past. \citet{Volger97},
studying {\it ROSAT} PSPC and HRI detected X-ray sources in NGC 4449,
found an X-ray (0.1-2.4 keV) luminosity of 4.7 $\times$ 10$^{38}$ erg
s$^{-1}$, a value approximately 50\% of the value quoted by
\citet{Blair83} using data from the {\it Einstein   X-ray Observatory}
High-Resolution Imager. This apparent decline in emission agreed with
more recent observations taken with the {\it   Chandra X-ray
Observatory} which found L$_{x}$ = 2.4 -- 4.0 $\times$ 10$^{38}$ erg
s$^{-1}$ \citep{Patnaude03,Summers03}. Similarly, a marked decline in
emission at all radio frequencies has been observed with a drop of 13
mJy to 4 mJy from 1973 to 2002 at 4.9 GHz \citep{Lacey07}.  

The underlying cause of SNR 4449-1's extraordinarily high luminosity
has been proposed to be due to an especially strong interaction of the
remnant with dense local interstellar and/or circumstellar material
(ISM/CSM). \citet{Kirshner80} argued that the nature of the high
luminosity could be best explained if the remnant was in fact embedded
inside an \ion{H}{2} region, and \citet{Blair83} favored the scenario
of the SNR expanding into an ordinary  \ion{H}{2} region ($n
_{0}\approx$ 25 cm$^{-3}$) as opposed to a much denser ($n
_{0}\approx$ 150 cm$^{-3}$) medium. Later observations of the object
taken in the optical \citep{Bowmans97} have supported the general
notion of strong interaction with an \ion{H}{2} region.    

Here we present optical ground-based and {\it HST} images and spectra
that suggest the remnant may be interacting, not with an H II region,
but rather with dense circumstellar material from a massive progenitor
star.  In \S \ 2 we describe observations of the object with 
results from the images and spectra presented in \S \ 3.  We then
discuss the nature of SNR 4449-1's high luminosity, describe its
optical evolution, and compare it with other remnants and late-time
SNe observations in \S \ 4. 

\section{Observations}

\subsection{Images}

A 200 s exposure of NGC 4449 was taken at MDM Observatory on 1996 May
21 with the 1.3 m McGraw-Hill telescope on Kitt Peak and a 2048
$\times$ 2048 pixel CCD with a plate scale of 0$\farcs$44 pixel$^{-1}$
yielding a 14$\farcm$9 $\times$ 14$\farcm$9 field of view. A portion
of this image is shown in Figure 1. A filter centered at $\lambda$ =
5000 \AA  \ (FWHM = 110 \AA) was used to detect SNR~4449-1's strong
[\ion{O}{3}] $\lambda\lambda$4959,5007 line emission and place the
remnant in context to neighboring H~II regions.  

High resolution images of NGC~4449 were recently obtained by the
Advanced Camera for Surveys (ACS; \citealt{Ford98,Pavlovsky04}) system
onboard {\it HST} using the Wide Field Channel (WFC), some of which
covered the SNR~4449-1 region.  A portion of some of these images were
kindly made available before their public release by D. Calzetti, with
the rest retrieved from the STScI archives.  The ACS/WFC consists of
two 2048 $\times$ 4096 pixel CCDs providing a 202$''$ $\times$ 202$''$
field of view with an average pixel size of 0$\farcs$05. Standard
pipeline IRAF/STSDAS data reduction was done including debiasing,
flat-fielding, geometric distortion corrections, photometric
calibrations, and cosmic-ray and hot pixel removal, and the STSDAS
\texttt{drizzle} task was used to combine exposures in each filter.  

Table 1 lists ACS images of NGC~4449 that covered the O-rich SNR and
were examined.  The F502N filter is sensitive to the [\ion{O}{3}]
$\lambda\lambda$4959,5007 lines, whereas F550M is a line-free
continuum filter.  The F658N filter passes both H$\alpha$
$\lambda$6563 and [\ion{N}{2}] $\lambda\lambda$6548,6583 emission
lines, while F660N isolates the [\ion{N}{2}] $\lambda$6583 emission
line. The broadband filters F435W, F555W, and F814W are equivalent to
Johnson {\sl B}, {\sl V}, and {\sl I} filters, respectively.    

We used the ACS module of the DOLPHOT software \citep{Dolphin00} to
obtain  photometry on these images.  DOLPHOT is a point-spread
function fitting package specifically tailored to do photometry of
{\it HST} images. Photometry was done on the flat-fielded images from
the STScI archive using the drizzled F435W as the reference image. The
package identifies the sources and performs the photometry on
individual frames, also taking into account all the information about
image cosmetics and cosmic-ray hits that is attached to the
observational material. DOLPHOT provides magnitudes corrected for
charge-transfer efficiency effects on the calibrated VEGAMAG scale
described by \citet{Sirianni05}. The package yields a variety of
information for each object detected, including the object type
(stellar, extended, etc.), $\chi ^2$ of the PSF fit, sharpness and
roundness of the object, as well as a ``crowding'' parameter that
measures how much brighter an object would have been had neighboring
objects not been fit simultaneously. We selected objects of stellar
type, valid photometry, $\chi ^2$ $<$ 5, and sharpness between $-0.35$
and $+0.35$.    

We also retrieved from the STScI archives high resolution images of the
remnant taken on 1996 February 22 by the Faint Object Camera
previously onboard the {\it  HST}.  The images have a field of view of
7$\farcs$3 $\times$ 7$\farcs$3 (512 $\times$ 512 pixels, with each
pixel 0$\farcs$014 $\times$ 0$\farcs$014 in size).  One image of
exposure length 1635 s was taken with the F501N filter sensitive to
[\ion{O}{3}] ($\lambda_c$ = 5010 \AA; FWHM = 74 \AA), and
another of exposure length 596 s taken with the medium band F346M
filter ($\lambda_c$ = 3480 \AA; FWHM = 434 \AA).  These
COSTAR-corrected images were reduced in the standard STScI routine
science data pipeline that performed flat-fielding and geometric
corrections.    

\subsection{Spectra}

Low-dispersion, optical spectra of SNR~4449-1 were obtained at MDM
Observatory on Kitt Peak using the 2.4 m Hiltner telescope with a
Modular Spectrograph and 2048 $\times$ 2048 pixel SITe CCD detector.
Spectra were taken on 2002 June 17 and 2006 June 19 with total
integration times of 2160 s and 2700 s, respectively.  Seeing was
around 1$''$ (FWHM) for the 2002 spectra and 1.5$''$ -- 2.0$''$ in
2006.  A N--S 1.0$''$$\times$ 5.0$'$ slit and a 600 line mm$^{-1}$
5000 \AA \ blaze grism was used to obtain exposures spanning the
spectral region 4300 -- 7500 \AA \ with a resolution of 4 \AA.   

The remnant was observed again at MDM Observatory using the same
telescope, CCD detector, and spectrograph on 2007 April 9 -- 15.  The
600 line mm$^{-1}$ 5000 \AA \ blaze grism was used but now centered at
4500 \AA \ to improve sensitivity in the blue and expand our lower
spectral range to 4000 \AA.  We also used an 830 line mm$^{-1}$ 8465
\AA \ blaze grism centered around 7000 \AA \ for a spectral region
spanning 6100 -- 8500 \AA \ and a resolution of 3 \AA.  All spectra
were reduced using standard IRAF\footnote{The Image Reduction and
Analysis Facility is distributed by the National Optical Astronomy
Observatories, which are operated by the Association of Universities
for Research in  Astronomy, Inc., under cooperative agreement with
the  National  Science Foundation} routines and calibrated with Hg,
Ne, and Xe lamps and standard stars.   

Archival optical spectra of the remnant taken with the Faint Object
Spectrograph (FOS) aboard the {\it HST} were also retrieved from STScI and
examined (see \citealt{Blair96}).  These spectra were taken on 1993 Jan 28
(pre-COSTAR correction) using a circular 1$''$ diameter aperture in combination
with two gratings: (1) the G400H in FOS/RD configuration yielding spectral
range 3400 -- 4600 \AA \ with resolution 4.0 \AA, and (2) the G570H in FOS/RD
configuration with spectral range 4800 -- 6600 \AA \ and resolution 5.9 \AA.
The total exposure time was 1250 s for each spectrum. The quality comment
states that target acquisition failed during the pointing, so precise placement
of the remnant inside the aperture is uncertain. Data were processed through
standard FOS pipeline calibration, which includes detector background and
scattered light subtraction, flat-field corrections, computation of
wavelengths, and conversion from count rates to fluxes.  

Measurements of observed flux reported from the spectra and photometry  
are corrected for extinction with A$_{V}$ = 0.7 mag found
from the H$\alpha$/H$\beta$ ratio of our spectra. Reddening
corrections were applied using this value and
following the \citet{Cardelli89} reddening law with R$_{V}$ = 3.1.   

\section{Results}

\subsection{Environment Around the SNR}

The high resolution ACS/WFC images covering the SNR~4449-1 region
provide the first clear look at its stellar and circumstellar
environment (see Fig.~2).  In the F502N [\ion{O}{3}] image  (Fig.~2a),
one sees that the remnant is situated well to the east of two bright
\ion{H}{2} regions.  A high contrast version of this same image and
covering the same region is shown in Figure 2b, illustrating that the
remnant is by far the brightest unresolved source in this area at 5000
\AA.  The presence of a small possible faint [\ion{O}{3}] emission
ring surrounding the remnant will be discussed below in \S \ 3.3. 

The ACS/WFC H$\alpha$-sensitive F658N image presented in Figure 2c
places the remnant most clearly in context with respect to the local
\ion{H}{2} regions. Previous ground-based observations had suggested
the remnant to lie in or at the edge of the \ion{H}{2} regions located
immediately to the west (e.g. \citealt{Bowmans97}). However, the high
resolution {\it HST} image shows the remnant to lie well away from the
strong H$\alpha$ emission associated with these regions. The image
further reveals the SNR to be situated inside the cavity of an
irregularly shaped emission shell approximately 8$''$ $\times$ 6$''$
(150 $\times$ 110 pc) in size.  This shell is only weakly visible in
[\ion{O}{3}] (Fig.~2a) and barely noticeable in the [\ion{N}{2}]
(F660N) filter (Fig.~2d).  

The location of SNR 4449-1 and the large H$\alpha$ shell with respect
to the local stellar environment is shown in Figure 2e and Figure 2f
via the F435W and F550M continuum images.  These images reveal the
remnant to be situated inside a cluster of blue stars approximately
3$''$ $ \times$ 4$''$ in size, all of which lies within the emission
shell seen in the H$\alpha$ + [\ion{N}{2}] (F658N) image. 

Photometry of this stellar cluster near SNR~4449-1 was conducted using
the F435W, F555W, and F814W images.  Only stars that were observed to
be projected inside the H$\alpha$ + [\ion{N}{2}] emission shell and
within approximately 3$''$ of the remnant were measured.  Figure 3 is
a color-magnitude diagram of the resulting photometry, overlayed with
isochrones shifted to the distance, extinction and reddening of the
remnant to determine the most likely age of the cluster.  The Geneva
Group non-rotating stellar evolution models of \citet{Charbonnel93} are
shown with metallicity Z = 0.004 to best match the estimate of Z =
0.005 for NGC~4449 made by \citet{Annibali07}.     

Following \citet{Massey00}, we assume that the stars of this
association evolve in coeval fashion and use the turn-off mass of the
cluster to make an estimate of the mass of the progenitor star of the
remnant. As gauged by the cut-off of the youngest stars in the cluster,
the lower limit of the age of the stellar population is approximately
log (age yr$^{-1}$) = 7.25 $\pm$0.50.  This is an age consistent with
the expected lifetime of stars with M$_{\rm ZAMS}$ $\simeq$ 15 -- 25
M$_{\odot}$.

\subsection{The SNR}

The O-rich SNR is bright and unresolved in all of the narrowband {\sl
HST} images.  The FWHM size of the remnant in the [\ion{O}{3}]
(F502N), H$\alpha$ + [\ion{N}{2}] (F658N), and  [\ion{N}{2}] (F660N)
images is approximately 2.3 pixels (0$\farcs$12), comparable to the
FWHM of stars on the images.  The remnant has apparent magnitudes of
m$_{\rm F502N}$ = 16.274 $\pm$0.002, m$_{\rm F658N}$ = 18.371
$\pm$0.005, and m$_{\rm F660N}$ = 18.435 $\pm$0.008.  Previous
spectra found little indication of any hydrogen and nitrogen
emissions originating from the remnant (e.g. \citealt{Kirshner80},
\citealt{Blair83}), so we found it surprisingly bright in the
H$\alpha$ + [\ion{N}{2}] and [\ion{N}{2}] 6583 images. 

To distinguish between H$\alpha$ and [\ion{N}{2}] 6583 emissions from
the remnant, we followed the absolute calibration for the F550M,
F658N, and F660N filters outlined by \citet{ODell04}. The F658N filter
is nearly equally as good at transmitting both the H$\alpha$ and
stronger [\ion{N}{2}] 6583 lines, and the calibration procedure
provides equations to separate their individual emission line fluxes.
Using the count rates obtained with the DOLPHOT software, and
correcting the calibration constants of the F658N and F660N filters
for the radial velocity 207 \kms \ of NGC 4449 \citep{Schneider92}, we
estimate a [\ion{N}{2}] 6583/H$\alpha$ ratio of $\approx$ 1.    

Bright, continuous emission at the location of the remnant is also
observed in all of the wideband ACS/WFC images in the F435W, F555W,
and F814W filters. While these filters are sensitive to various
emission lines of the remnant, the F550M filter, with a bandwidth
range between 5200 -- 6000 \AA, is not, making it only sensitive to
stellar continuum.  The DOLPHOT software identifies three overlapping
sources at the precise location of the remnant in the F550M image: Two
sources to the northwest and southwest of the remnant with magnitudes
m$_{\rm   F550M}$ = 23.120 $\pm$0.025 and 23.390 $\pm$0.028,
respectively, and another with magnitude m$_{\rm F550M}$ = 22.246
$\pm$0.016 almost coincident  (see Fig.~4a and 4b). Using the
\texttt{SYNPHOT} package and the Bruzual Spectrum Synthesis Atlas to
simulate the expected count rate from a stellar source, the entire
flux within a 0$\farcs$4 aperture around the center of emission
containing all three sources in the F550M image is equivalent to
approximately five O5V stars ($\sim$ 3.4 $\times$ 10$^{-15}$ erg
s$^{-1}$ cm$^{-2}$).      

Aligning the ACS/WFC images with the higher resolution FOC image adds
clarity with respect to the three stellar sources identified by
DOLPHOT at the remnant's location. Employing the \texttt{geomap} and
\texttt{geotran} routines in IRAF, the FOC F346M image was aligned
with the ACS/WFC F502N and F550M images (see Fig.~4c).  While the
alignment is prone to some uncertainty because of differences in scale
and geometric distortion between the two instrument configurations,
the placement of the remnant is quite clear. The SNR as defined by its
[\ion{O}{3}] emission coincides almost exactly with the brightest
source of F346M emission, with smaller, less bright sources located to
the northwest and southwest.  These sources are labeled 1 -- 3,
respectively, in Figure 4c. Furthermore, the alignment shows that the
locations of the three sources observed in the F346M image matches the
arrangement of the continuum sources identified in the F550M image.

\subsection{A Circumstellar Ring?}

An apparent faint extended ring of emission situated immediately
around the SNR, approximately 1$''$ in diameter ($\sim$ 20 pc), is
visible in the [\ion{O}{3}] (F502N) image.  To investigate the reality
of this feature, the [\ion{O}{3}] (F502N), H$\alpha$ + [\ion{N}{2}]
(F658N), and [\ion{N}{2}] 6583 (F660N) images were
continuum-subtracted using the line-free F550M image (see Fig.~5).  As
shown in this figure, the 1$''$ ring is still weakly detected in
[\ion{O}{3}] after the subtraction, with a strength 0.3\% of the peak
flux of the remnant and $\sim$ 3 times above the background
signal. The ring is not detected, however, in either H$\alpha$ +
[\ion{N}{2}] or [\ion{N}{2}] 6583.    

Initially, some properties of this apparent [\ion{O}{3}] emission ring
supported  its authenticity.  The ring structure appears clumpy and
elliptical, features not normally associated with Airy diffraction
rings or artifacts of filter reflections or scattering. However, a
bright field star located approximately 45$''$ east of the center of
the galaxy in the F502N image was found to possess a ring-like
structure sharing many similarities with that seen around
SNR~4449-1. When overlayed with one another, the rings are of equal
size (width  $\approx$ 0$\farcs$1) and have the same relative position
from the center of emission. There is also a similar broken pattern in
the encompassing ring, with a correspondence between the breaks in the
ring of the field star and the apparent clumps in the thin ring around
the remnant.  Additional archival ACS/WFC images  taken with the F502N
filter were also retrieved to further investigate the nature of the
[\ion{O}{3}] ring, and we again found the same broken ring of equal
size surrounding the brightest objects.  
  
On the other hand, the correspondence between the ring features of the
remnant and other bright sources is not perfect. Thin emission rings
around bright field stars are perfectly round while the ring around
SNR~4449-1 is noticeably elliptical with the major axis nearly N--S.
Specifically, faint emission toward the north--northwest does not
correspond to any features seen around field stars. Consequently, we
conclude that much of the apparent
[\ion{O}{3}] emission ring seen in the F502N image is a diffraction
ring, but with the possibility of some real [\ion{O}{3}] emission
surrounding the remnant especially to the northwest.       
 
\subsection{Optical Spectra}

\subsubsection{Temporal Change in Spectra}

A low-dispersion optical spectrum of the remnant taken in 2002 at MDM
Observatory is presented in Figure 6. The spectrum shows the same two
components previously observed by \citet{Kirshner80}: (1) narrow lines of
H$\alpha$, H$\beta$, [\ion{N}{2}] and [\ion{S}{2}] associated with
\ion{H}{2} region-like emission, and (2) broad lines of [\ion{O}{1}],
[\ion{O}{2}] and [\ion{O}{3}] associated with ejecta of a young,
O-rich SNR.  The forbidden oxygen expansion velocity $V_{\rm exp}$
$\sim$ 6000 km s$^{-1}$, measured from the half width at zero
intensity from the [\ion{O}{3}] 5007 emission line
toward the red, agrees with the value reported by \citet{Blair98}.  

However, this 2002 spectrum also reveals emergent features not
observed in the 1978 -- 80 optical spectra of the SNR taken by
\citet{Kirshner80} and \citet{Blair83}. Broad, blueshifted emission
lines with similar velocity profiles of Si-group elements [\ion{S}{2}]
6716,6731, [\ion{Ar}{3}] 7136 and [\ion{Ca}{2}] 7291,7324 are visible.
Measured toward the blue and red of the [\ion{S}{2}] 6716 and 6731
lines, respectively, the velocities are estimated to lie in the range
of $-$2500 $\leq$ $V_{\rm exp}$ $\leq$ 400 \kms. The reduced and
background-subtracted 2D spectrum shown in Figure 7 illustrates the
blueshifted [\ion{S}{2}] and [\ion{Ar}{3}] emission features more
clearly than the 1D spectrum presented in Figure 6.  The [\ion{Ca}{2}]
emission, on the other hand, is difficult to see in both 1D and 2D
spectra because it is blended with the strong [\ion{O}{2}] 7320,7330
lines.

\subsubsection{Broad H$\alpha$ + [\ion{N}{2}] emission}

Previous optical spectra noted a lack of hydrogen from the remnant
\citep{Kirshner80}.  However, our ground-based spectra reveal the
presence of faint, broad emission extending from around 6540 to 6605
\AA \ centered near the narrow H$\alpha$ line.  Analysis of the raw
spectrum suggests the observed extended emission is most likely the
combination of slightly broadened [\ion{N}{2}] 6548,6583 and
H$\alpha$, each with velocities around 500 km s$^{-1}$ (see Fig.~7).   

Additional evidence for the presence of broad H$\alpha$ emission comes
from archival FOS spectra of the remnant, which we present in Figure
8. Our ground-based spectra in the optical lack the spatial resolution
to completely isolate the remnant from the \ion{H}{2} regions that lie
less than 2$''$ away. The {\it HST} observations, however, make it
possible to observe the remnant independent of the \ion{H}{2} regions'
strong emission lines that contaminate the spectra.  This advantage
makes the FOS spectra more sensitive to any faint H$\alpha$ and
[\ion{N}{2}] emission that can be attributed to the remnant.

Similar to our ground-based observations, the FOS spectrum also shows an
extended base in the region of H$\alpha$ spanning 6540 -- 6605 \AA. The
H$\alpha$ and [\ion{N}{2}] 6548,6583 lines are noticeably
broad, so much so that the [\ion{N}{2}] 6548 emission line is barely
discernible underneath the broad H$\alpha$.  Deblending the lines, the
profiles show $V_{\rm exp}$ = 500 $\pm$100 km s$^{-1}$ for the
[\ion{N}{2}] and H$\alpha$ lines, with the FWHM of the nitrogen lines
appearing somewhat larger than H$\alpha$. The [\ion{N}{2}]
6583/H$\alpha$ $\approx$ 0.5 ratio is slightly lower than our derived
ratio from the ACS/WFC images, which is a discrepancy likely due to
contamination from the surrounding diffuse \ion{H}{2} emission
(observed in the F658N image; see Fig.~2c) in the spectra that biases the
H$\alpha$ flux. 

\subsubsection{Other Emission Lines}

In the near UV of the FOS spectrum, the [\ion{Ne}{3}] 3869 and
[\ion{O}{2}] 3727 lines have similar velocity profiles with a maximum
expansion velocity of $\sim$ 5000 km s$^{-1}$.  In Figure 9 we show
the velocity line profiles of [\ion{O}{2}] 3727, [\ion{Ne}{3}] 3869
and [\ion{O}{3}] 4959,5007 that have been smoothed with a five pixel
boxcar and scaled in arbitrary units of flux. The profiles share
conspicuous emission peaks symmetrically spaced at approximately
$\pm$1600 \kms.  These minor emission peaks are present in both the
{\sl HST} FOS and ground-based spectra. Moreover, the emission
profiles of [\ion{O}{1}] 6300,6364 and [\ion{O}{2}] 7320,7330 of the
ground-based spectra share symmetric emission peaks around
$\pm$1600 \kms. Interestingly,  hints of minor emission peaks in the
broad [\ion{S}{2}] and [\ion{Ar}{3}] lines are also observed at
$-$1600 \kms. 

The emission line profile of [\ion{O}{3}] 4959,5007 seen in Figure 9
shows a staircase shape slightly skewed toward the blue.  The peaks
marked A and B correspond with emission from the lines at 4959 and
5007 \AA, respectively. We interpret the adjacent emission peaks
marked A$_{\rm   b}$ and A$_{\rm r}$ as [\ion{O}{3}] 5007 emission
blueshifted and redshifted by $\pm$1600 \kms. Likewise, we interpret
the minor emission peaks marked B$_{\rm b}$ and B$_{\rm   r}$ as
[\ion{O}{3}] 4959 emission similarly blueshifted and redshifted.  The
overlap between the 4959 and 5007 emission lines heightens the blue
side of the profile.  

Hints of two minor emission peaks at +3000 and +4600 \kms \ are
also seen in the [\ion{O}{3}] profile of the FOS spectra. The ground-based
2002 -- 2007 spectra show some faint evidence for these peaks as
well. However, they do not correspond with any distinguishable
emission observed within the [\ion{O}{2}] or [\ion{Ne}{3}] lines.   We
compared the [\ion{O}{3}] 4959,5007 emission peaks at +3000 and +4600
\kms \ of the FOS spectra with the [\ion{O}{1}] 6300,64 and
[\ion{O}{2}] 7320,30 lines of the ground-based spectra and did
not find significant similiarity.   

Additional faint emission in the form of a clump is observed in the forbidden
oxygen lines of the ground-based spectra at velocities between $-$3000 and
$-$6000 \kms.  In Figure 7 we highlight an example of this emission (marked ``O
Clump''), seen blueward of the [\ion{O}{2}] 7320 line with a minor peak around
$-$4600 \kms.  Similarly blueshifted and faint emission is observed in the
[\ion{O}{3}] 4959 line where a noticeable bump in emission from the remnant
merges into the H$\beta$ line of the nearby \ion{H}{2} region (see Fig. \ 6).
Though the clump is less pronounced in the remnant's [\ion{O}{1}] 6300 line,
there is a noticeable blueshifted ledge of emission beginning around $-$3000
\kms \ and extending out to $-$6000 \kms.   

The FOS spectra also weakly detects the temperature-sensitive emission
line at [\ion{N}{2}] 5755.  The relatively low [\ion{N}{2}]
(6548+6583)/5755 $\la$ 15 ratio implies unreasonably high temperatures
of $\sim$ 10$^{5}$ K for densities between 100 -- 1000 cm$^{-3}$, so
we suspect collisional de-excitation is quenching the [\ion{N}{2}]
6548,83 lines.  For this effect to be observed, electron densities
greater than log n$_{\rm   e}$ = 4.8 are required. If we assume a
temperature in the region of T $\sim$ 10$^{4}$ K, the electron density
is in the range 1 -- 1.5$\times$10$^5$ cm$^{-3}$.

\subsubsection{A Wolf-Rayet Signature?}

In our ground-based spectra, a weak line is observed near 4689 \AA \
that we suspect is \ion{He}{2} 4686, and it is coincident with broad,
faint excess emission approximately 100 \AA \ wide extending between
4600 -- 4700 \AA \ (see inset of Fig.~7).  The detection of this
feature is slight (S/N $\sim$ 3), but its placement and shape is
similar to a ``WR bump'' indicative of the presence of Wolf-Rayet (WR)
stars at the location of the remnant. The FOS spectra show the same
faint excess emission around the \ion{He}{2} 4686 emission line,
supporting our suspicion. Additional helium lines of \ion{He}{1} at
5876, 6678, and 7065 are observed in the MDM spectra, but these
originate from the nearby \ion{H}{2} region and not from the remnant
itself.     

\section{Discussion}

\subsection{Origin of SNR~4449-1's Extraordinary Brightness}

Young SNRs such as SNR 4449-1 are associated with two shocks likely to
produce heating and observable radiation: (1) an outward-propagating
blast wave running into surrounding material (CSM and/or ISM) giving
rise to bright radio flux, and (2) an inward-propagating reverse shock
that heats supernova ejecta and emits strongly across a wide spectral
band. Previous studies of SNR 4449-1 attributed its high luminosity to
a strong interaction between the remnant and a dense \ion{H}{2}
environment believed to encompass it \citep{Kirshner80,Blair83}.
However, the high resolution ACS/WFC images clearly show the remnant
isolated from the \ion{H}{2} regions some 2$''$ away to the west (see
Fig.~2c).  

As assumed by previous analyses, we attribute the extremely broad ($V_{\rm
exp}$ $>$ 1000 km s$^{-1}$) components of the spectra to interaction between a
reverse shock and the expanding O-rich debris. However, instead of the reverse
shock originating from the blast wave running against an \ion{H}{2} region
(n$_{\rm e}$ $\approx$ 25 cm$^{-3}$), the observations presented above strongly
suggest that SNR 4449-1 is interacting with very dense and extensive
circumstellar material. The intermediate velocity (500 $\pm$100 km s$^{-1}$)
H$\alpha$ and [\ion{N}{2}] emissions  likely originate from shock-heated CSM,
and the high [\ion{N}{2}] 6583/H$\alpha$ ratio $\approx$ 1 indicates it is
N-rich. If we adopt the post-shock density of n$_{\rm e}$ $\sim$ $10^5$
cm$^{-3}$ implied by the [\ion{N}{2}] 5755 emission line and assume a
temperature T $\sim$ 10$^4$ K, the measured H$\alpha$ flux of 1.0 $\times$
10$^{-14}$ erg s$^{-1}$ cm$^{-2}$ suggests a CSM mass of $\sim$ 0.5
M$_{\odot}$.   

Alternatively, given that the remnant is coincident with bright
stellar sources (see Fig.~4), it is also possible that UV emissions
and/or interaction between the remnant and the stellar winds of nearby
stars could contribute to the observed luminosity.  The aligned FOC
and ACS images show two stellar sources to the northwest
and southwest of the remnant little more than $1$ pc away, while the
center of the relatively bright continuum source observed in the
line-free (F550M) image lies within $\la$ 0.3 pc of the center of the
remnant as observed in the [\ion{O}{3}] (F502N) image. In light of the
high density inferred for the CSM, we view it as more likely that the
continued brightness of SNR 4449-1 to be the result of strong, ongoing
interactions between the remnant and an extensive circumstellar
environment left behind by its massive progenitor.  However, the fact
that the remnant is located within a compact stellar grouping less
than a few parsecs in size allows for the possibility that it could
also be interacting with the wind material of closely neighboring
stars.    

\subsection{The Progenitor Star}

If the assumption of coeval evolution is appropriate for the stellar
cluster SNR 4449-1 is located in, the mass range of 15 -- 25
M$_{\odot}$ derived from the photometry and stellar evolutionary
tracks then represents a minimum mass for the progenitor. Our photometry
and isochrone fitting agree with the results of \citet{Annibali07}
who used the same {\it HST} images but in combination with Padova
stellar evolutionary tracks \citep{Fagotto94a,Fagotto94b}, and our
mass estimate derived from the turn-off mass of the cluster is consistent
with \citet{Blair83} who argued for a progenitor star mass of $\sim$ 25
M$_{\odot}$ by comparing elemental abundances observed from optical
and X-ray data with the stellar models of \citet{Weaver80}.

However, a 15 -- 25 M$_{\odot}$ mass range is likely a modest estimate
for the progenitor star. Because the remnant's emission lines extend
across the wide passbands of the F435W, F555W, and F814W filters, we
were unable to do photometry of the stars directly coincident with the
remnant.  Being located at the densest part of the cluster, these
stars may represent the most massive members, and, if so, could be of
larger mass than the remainder we were able to do photometry on.  Our
mass estimate would also be larger if stars in this cluster are
rotating.  The low metallicity models of \citet{Maeder01} show that
the lifetimes of massive rotating stars are longer than
their non-rotating counterparts, meaning masses on rotating isochrones
would be higher than the masses on non-rotating isochrones of the same
age. Thus, the progenitor star of SNR~4449-1 could very well be
greater than 20 M$_{\odot}$.   

Excess broad emission around 4600 -- 4700 \AA \ near the \ion{He}{2}
4686 emission line (see Fig.~8) is a well known feature of galaxies
and clusters of stars harboring a significant population of Wolf-Rayet
stars. Generally, this ``WR bump'' is a combination of many 
emission lines including \ion{N}{3} 4634,4642, \ion{N}{5} 4605,4622,
and \ion{C}{3}/\ion{C}{4} 4650,4658, in addition to \ion{He}{2}
4686. The presence of the WR bump in the spectra strongly suggests
that WR stars are present near the location of the remnant, and the
absence of carbon lines at \ion{C}{3} 5696 and \ion{C}{4} 5801,5812
suggests that most of the WR stars are of the WN type. Assuming the
progenitor star to be among the most massive of the stars located
here, the progenitor of SNR~4449-1 may have been a WR star itself.  

On the other hand, the relatively broad profiles of H$\alpha$ and
[\ion{N}{2}] with $V_{\rm exp}$ $\approx$ 500 \kms, along with the
high density $\ga$ 10$^{5}$ cm$^{-3}$ of the CSM implied by the
[\ion{N}{2}] 5755 line, are not consistent with properties observed
for WR stars and their circumstellar environments. WR stars are
typically associated with wind velocities $\ga$ 1000 \kms, and their
nebulae generally have densities $<$ 1000 cm$^{-3}$
\citep{Esteban92}. Instead, these observations are more in line with
properties found in luminous blue variables (LBVs) and their
circumstellar nebulae.  LBVs have slower wind velocities between 40 --
700 \kms, and higher densities in their circumstellar nebulae that,
due to their clumpy nature, range between 10$^{3}$ to 10$^{7}$
cm$^{-3}$ \citep{Stahl89}.    

An LBV/WR progenitor star of SNR~4449-1 is an interesting possibility.
However, such an implication is complicated by the inferred $\ga$ 20
M$_{\odot}$ progenitor star mass which just meets the threshold for
the mass of an O star to have mass-loss substantial enough to strip
away its outer atmosphere and develop into a WR star.  Furthermore,
LBVs are generally associated with stars of mass about 40 M$_{\odot}$
and greater. There are instances, however, where these evolved stars
have been observed to lie in the 20 M$_{\odot}$ mass range.  For
example, the well-studied WR stars of the Ofpe/WN9 type along with
their stellar and circumstellar environments share many of the
characteristics observed around SNR~4449-1. These stars are observed
to have similar mass ranges as derived from turn-off masses in stellar
associations \citep{Massey00}, are known to have a solid observational
link with LBVs \citep{Pasquali97}, and are frequently found in tight
($\la$ 1 pc) groupings \citep{Lortet89}.  

Aside from a single star scenario, a progenitor with a binary
companion is also a plausible way for a 20 M$_{\odot}$ star to evolve
into a WR star which in turn could lead to a SNe yielding an O-rich
SNR. We note that low metallicity environments like NGC~4449 are
understood to possess WR populations dominated by WR stars born in
high mass, close binary systems \citep{Eldridge06}.  Furthermore,
possible periodic variations in the overall downward decline in the
radio flux of SNR~4449-1 (see Fig. \ 2 of \citealt{Lacey07}) resemble
the radio fluctuations observed in SN~1979C believed to most likely be
consequence of interaction with a massive companion star in a highly
eccentric orbit \citep{Weiler92}. Thus, finding SNR 4449-1 in a tight
grouping of luminous stellar sources makes binarity a strong possibility.       

\subsection{SNR Evolution}

The optical spectra of SNR 4449-1 shows a number of emergent features
when compared against spectra originally obtained in 1978 -- 80. The
most obvious change that we observe is the development of broad,
blueshifted emissions from [\ion{S}{2}], [\ion{Ar}{3}], and
[\ion{Ca}{2}].  The only previous evidence for broadening outside of
forbidden oxygen was in the [\ion{S}{2}] 4069,4076 lines, first
observed by \citet{Balick78} and later better resolved by
\citet{Blair83}.  \citet{Kirshner80} could find no evidence of broad
[\ion{S}{2}] 6716,6731 lines, making the presented  spectra the
first observation of broadening around these lines, and for all
of the Si-group elements in general.   

Evolution of the optical spectra of this sort had been anticipated in
previous studies \citep{Balick78,Seaquist78,Kirshner80}. The new line
emission detections likely arise from the reverse-shock heating of a
portion of SN ejecta (possibly a plume) of mixed O- and S-Ar-Ca-rich
ejecta material.  Why these blueshifted emissions were not observed
in previous spectra and are only observable now appears to be due in
part to the continuing expansion of the ejecta, where the densities of
the SN debris drop below the critical densities for collisional
de-excitation of various S-Ar-Ca lines. 

From an examination of the optical spectra of SNR 4449-1 in \citet{Blair83}, we
estimate the broadening in the [\ion{S}{2}] 4069,4076 emission lines to
have FWZI $\approx$ 3500 \kms. This width corresponds with the broadening 
in the [\ion{S}{2}] 6716,6731 lines measured in the 2002 spectra.  The
[\ion{S}{2}] emission lines have a critical density of log n$_{\rm e}$
$\approx$ 3.4, which is lower than the log n$_{\rm e}$ $\approx$ 6.4 critical
density of the [\ion{S}{2}] 4069,4076 lines. Hence, it is a natural
consequence for the the red [\ion{S}{2}] 6716,6731 lines to be visible
following a decline in the density of the expanding
shock-excited ejecta.    

Finally, we note that minor emission peaks in the velocity profile of the
forbidden oxygen lines of SNR 4449-1, first recognized by
\citet{Kirshner80} and seen in our more recent spectra, suggest 
the presence of clumpy ejecta with a possible symmetry between
facing and rear expanding hemispheres.  The matching minor
peaks in the [\ion{O}{2}], [\ion{Ne}{3}], and [\ion{O}{3}] emission
line profiles suggest clumping in the ejecta at velocities around $\pm$1600
\kms, and the symmetry of the emission suggests a possible ring or jet
distribution.  Moreover, observing minor emission peaks
in the [\ion{S}{2}] and [\ion{Ar}{3}] profiles around $-$1600 \kms \
suggests that these elements are constituents of the O-rich expanding
ejecta and share a similar spatial distribution. 

The origin of additional minor emission peaks in the [\ion{O}{3}] 5007
profile around +3000 and +4600 \kms \ is not known.  It is
curious that the peaks are symmetric with the [\ion{O}{3}] 4959
emission line at its rest and blueshifted ($-$1600 \kms) wavelengths.
Because these fainter, higher velocity peaks are not observed in
the [\ion{O}{2}] 3727 and [\ion{Ne}{3}] 3869 emission lines, their
reality is less certain than the stronger ones at $\pm$1600 \kms.
Spectra with better S/N of the [\ion{O}{1}], [\ion{O}{2}],
[\ion{O}{3}], and [\ion{Ne}{3}] lines would help investigate the
possibility of additional minor emission peaks at velocities higher
than 1600 \kms. 

\subsection{Size and age of the remnant}

Though SNR~4449-1 is clearly a young object, the lack of a reported SN
in NGC~4449 near the remnant's location leaves its precise age
uncertain. Assuming the remnant is still in a free expansion phase
despite its prodigious energy output ($\ga$ 10$^{49}$ erg across the
radio, optical, and X-ray), one can obtain an age estimate from
measurements of its angular size and expansion velocity.

The observed size of the remnant in the FOC F501N image is just
$\approx$ 4 pixels (FWHM), meaning it is marginally resolved above the
instrumental FWHM = 3 pixels (FOC Handbook v7.0, Fig.~26).  With a
0$\farcs$014 pixel$^{-1}$ ratio, this implies a remnant angular
diameter of $\approx$ 0$\farcs$037, which is slightly larger than the
value of 0$\farcs$028 estimated by \citet{Blair98}. Using the latest
distance  estimate to NGC 4449 of 3.82 $\pm$0.18 Mpc
\citep{Annibali07}, the maximum expansion velocity of 6000 km s$^{-1}$
for [\ion{O}{3}] implies a present age of $\sim$ 50 yr. This estimate
is limited by both the uncertainty in the distance to NGC 4449
(between 2.93 and 5 Mpc; [\citealt{Kara98,Sandage75}]) and the fact
that the remnant is just barely resolved above the instrumental
profile.  However, we can conclude that the remnant is probably not
much older than 100 yr since at ages above this, its size ($>$ 1.2 pc)
should have been more clearly resolved in the 1996 FOC image.  

Previous attempts at dating SNR~4449-1 have tried to place constraints
on its age through historical plate searches to find a serendipitous
detection of the associated SN outburst.  \citet{deBruyn81} examined
plates dating back to 1925 taken at Mt.\ Wilson Observatory and found
no change in the appearance of the SNR~4449-1 region. \citet{Blair83}
inspected the patrol plates of the Harvard College Observatory
including the RH series dating 1928 to 1954 with limiting magnitude
$m_{V}$ = 14 -- 15 and the AC series dating 1898 to 1954 with $m_{V}$
= 12. They too were unable to find any change in appearance at the
location of the remnant.  

We complemented the \citet{Blair83} plate search by examining a dozen
of the Harvard MC series plates from 1917 to 1954 with limiting magnitude
$m_{V}$ = 17 -- 18 and were also unable to find any observable change
in the location of the remnant. Assuming an absolute luminosity
$M_{V}$ = $-$17 for a typical core-collapse SN along with a $\sim$ 4
Mpc distance, the AC series of plates are barely within range of
detection of the original SN under ideal observing conditions. And
even though the MC and RH series plates are deep enough to image the
original SNe, the time coverage and/or unfortunate timing of the SN
outburst when NGC~4449 was behind and thus hidden by the Sun easily
accounts for why these searches are inherently ambiguous and could
easily miss the original supernova event.    

Possibly complicating matters, eariler plate searches may have
focused on the wrong time frame.  \citet{Kirshner80} and
\citet{Blair83} estimated the age of SNR~4449-1 to be $\sim 100 - 200$
yr using an expansion velocity of 3500 \kms.  In light of recent
observations, however, the velocity is really closer to 6000 \kms,
implying a younger remnant with an age in line with our $\sim 50 - 100$
yr estimation from the {\it HST} data.  The consequence is that the
prior searches, which were mostly limited to observations
made previous to 1954, may have missed the original supernova if it
did indeed occur later in time.   

In Figure 10 we present archival images of NGC~4449 that set a minimum
date for the original SNe.  Figure 10a is a plate taken by P.~Hodge in
1965 with the Lick 3 m telescope using 103aD emulsion behind a GG 11
filter (a $V$-band image), and Figure 10b is the same image as Figure
10a but zoomed in on the location of the remnant. Figure 10c is the
2005 ACS/WFC F555W ($V$) image that has been smoothed to match the
resolution of the 1965 plate. The remnant is marked and stellar
sources to the east and west are highlighted to show a basis of
comparison. The two plate images show the SNR~4449-1 region
significantly brighter in 1965 than in 2005, and we interpret this as
definitively showing the remnant to be at least 42 yr old.  Not
presented is another plate that we examined, obtained with the same
emulsion and filter combination and taken four years earlier in 1961
using the 100 inch telescope at Mt.~Wilson.  This 1961 $V$-band plate
image shows the remnant just as bright, and thus further constrains
the age of the remnant to be at least 46 yr.          

Unfortunately, some images of NGC~4449 taken prior to 1961
cannot date the remnant with certainty.  As the 1965 image
illustrates, plates taken in the visual band between 5000 -- 6000 \AA
\ are sensitive to the broad [\ion{O}{3}] 4959,5007 lines and show the
remnant unambiguously when compared to the recent 2005 images where we
can observe a clear decline in luminosity. However, searches of older
plate material using emulsions sensitive to the 3000 -- 5000 \AA \
wavelength region are less conclusive because the emulsions are
preferentially sensitive to the UV light of the remnant's associated
OB cluster.   

This problem is illustrated in Figures 10d and 10e.  Shown in Figure
10d is a plate taken on 1913 April 7 with the Mt.~Wilson 60 inch
telescope using a blue-sensitive emulsion exposed for five hours.
Figure 10e, on the other hand, is the 2005 ACS/WFC F435W ($B$) image
smoothed to match the resolution of the 1913 plate. Comparing the two,
there is some drop in luminosity between 1913 and 2005, but nowhere
near the difference observed between the $V$-band 1965 and 2005
images.  We examined blue-sensitive images taken in 1917, 1935, and
1952, and the location of the remnant retained roughly the same
relative brightness. Thus, while it is tempting to associate the point
source located in the 1913 image with SNR~4449-1, images taken in
blue-sensitive emulsion do not definitively identify the remnant
because one cannot be sure whether the emission is due to the remnant
or the bright, compact star cluster it is located in.         

\subsection{Comparison of SNR~4449-1 with other Young SNRs and Old SNe}

SNR~4449-1's age of $\sim 50 - 100$ yr places it in an unique position
between the youngest Galactic supernova remnants that are many
hundreds of years old (Cas A being the youngest at around 325~yr;
\citealt{Thorstensen01}), and
the eldest of extragalactic core-collapse supernovae observed a few
decades after outburst.  In this way SNR~4449-1 is an interesting
link between SNe and their remnants, making it worthwhile to compare
its properties against both types of objects.  

SNR~4449-1 belongs to a small class of young O-rich supernova
remnants. The eight known members of this SNR subclass are believed to
represent the remains of high-mass stars ($\ga$ 15 M$_{\odot}$) having
debris with high velocities ($>$ 1000 km s$^{-1}$) and elevated
abundances of oxygen and neon. Cas A is the prototypical example of
this O-rich class of SNRs.  Observed in Cas A are large proper motion,
high radial velocity ejecta called ``Fast-Moving Knots'' (FMKs;
$V_{\rm exp}$ = 4000 -- 6000 \kms) that show strong O, S, and Ar
lines, but no H, He, or N emissions.  These knots have been
interpreted as fragments of debris from the progenitor's mantle and
core.  Also observed are much slower moving knots called
``Quasi-Stationary Flocculi'' (QSFs; $V_{\rm exp}$ $\leq$ 400 \kms)
that show both hydrogen Balmer emission and strong [\ion{N}{2}] lines
consistent with them being CNO-processed CSM.   

Considering SNR~4449-1 shares many optical emission properties with
Cas A, they may have had similar progenitor stars. High mass
progenitor estimates of Cas A are in the range of 10 to 30 M$_{\odot}$
\citep{Fabian80,Jansen88,Vink98}. The elevated abundances of O-burning
products in Cas A's FMKs, and the slower moving He- and N-rich QSFs of
circumstellar material, suggest that its progenitor may have been a
WNL star that experienced substantial mass loss before exploding as a
Type Ib/c or Type IIb/L SNe  \citep{Fesen91,Garcia96,Vink96}. On the
other hand, lower masses between 15 -- 25 M$_{\odot}$ have been
suggested if instead the progenitor star interacted with a binary
companion that stripped the star of its outer layers \citep{Young06}.   

If we look at SNR~4449-1 as a late-time SNe, there are very few
extragalactic, core-collapse supernovae aged $\ga$ 25 yr from which
to draw comparisons. Most SNe tend to fade rapidly in the optical and
become nearly unobservable after two years. This situation makes the
high optical brightness of SNR~4449-1 in light of its age both
interesting and a challenge to categorize.    

The eldest and most notable examples of SNe optically recovered many
years after outburst include SN 1957D \citep{Long89}, SN 1970G
\citep{Fesen93}, SN 1979C and SN 1980K \citep{Fesen99}. With the
exception of SN 1957D, which was not observed near maximum, these SNe
were all of Type II-L.  Collectively all show high velocity ($\sim$
5000 \kms) H$\alpha$, [\ion{O}{1}] 6300,6364 and [\ion{O}{2}]
7320,7330 emission features at late time. Yet all exhibited a flux
at least $\sim$ 10 times less than SNR~4449-1.  It would seem that,
SNR~4449-1, with its much lower velocity H$\alpha$ emission, greater
brightness, and dominating [\ion{O}{3}] emission, may have little in
common with these remnants of Type II-L.     

On the other hand, SNR~4449-1 does share features in its optical
spectra with two other notable recoveries of evolved SNe: SN 1986J
\citep{Leibundgut91} and SN 1978K \citep{Ryder93}. These SNe are both
of Type IIn, a class that has been associated with dense circumstellar
environments left behind by massive (possibly LBV) progenitor stars
(see, e.g. \citealt{GalYam07}). At several years of age, SN 1986J
showed intermediate-width $\Delta v \simeq$ 600 km s$^{-1}$ lines of
H, He, N and Fe, and broad $\Delta v >$ 1000 km s$^{-1}$ lines of
[\ion{O}{1}], [\ion{O}{2}] and [\ion{O}{3}]. Similarly, almost 20 yr
after outburst, SN 1978K possessed an intermediate-width (FWHM $\leq$
560 \kms) H$\alpha$ profile without a high velocity base. Aside from
sharing intermediate velocities, SNR~4449-1, SN 1986J, and SN 1978K
are among the rare instances where the [\ion{N}{2}] 5755 emission line
has been observed in evolved SNe \citep{Chu99}.

Some mention should also be made of SN 1987A, which at late time shows
features in its SN ejecta-CSM interactions like those we observe in
SNR~4449-1. SN 1987A was an SNe Type II-peculiar, and known to be the
descendant of a 20 M$_{\odot}$ blue supergiant progenitor. Currently
20 yr past outburst, SN 1987A is interacting with circumstellar
material along its equatorial ring that has, like SNR~4449-1, a high
[\ion{N}{2}] 6583/H$\alpha$ ratio and post-shock density as large as
$\sim$ 10$^{6}$ cm$^{-3}$ (see, e.g. \citealt{Pun02}).

\subsection{Conclusion}

We have presented a collection of images and spectra both ground-based
and taken by {\it HST} of the young O-rich supernova remnant in NGC
4449.  The nature of SNR~4449-1's bright luminosity was long suspected
to be due to interaction with a surrounding \ion{H}{2} region, but
new observations suggest that the remnant is instead interacting
with very dense N-rich circumstellar material from the SNR's
progenitor star of mass $\ga$ 20 M$_{\odot}$.  {\it{HST}} images show
that the remnant lies within a rich OB association and is located 
at the center of a tight grouping of bright blue stars less
than a few pc in size. 

Strong interaction of SN ejecta with a dense wind helps explain
SNR~4449-1's high luminosity, and its sustained brightness suggests an
extensive circumstellar environment. The advance of the forward shock
through this material gives rise to the broadened H$\alpha$ and
[\ion{N}{2}] emission seen, while the reverse shock from this
interaction heats the O-rich expanding ejecta that dominates the
presently observed optical spectrum. While we favor the scenario of
the remnant running into the CSM of its own progenitor, the close
proximity of the remnant to several bright stellar sources makes
interaction with the wind-loss material of other stars a possibility.  

We note that, despite its unusual luminosity, SNR~4449-1 may represent
a close match to a textbook high-mass progenitor SN scenario. That is
to say, SNR~4449-1 possesses the following characteristics thought to
be typical of SNe of massive stars: (1) the remnant lies within a rich
cluster of high-mass stars surrounded by a presumably SN- and
wind-blown ISM bubble, (2) it is located at the dense center of an
OB star cluster which appears to possess some WR stars, (3) it is
interacting with dense N-rich, circumstellar mass loss material, and
(4) it exhibits the expected chemical properties of an H-poor envelope
progenitor. Consequently, further studies of this bright SNR may help
clarify the future evolution of some recent high-mass SNe which also
show strong CSM interactions.

\acknowledgments

We thank John Thorstensen for obtaining the 2002 and 2006 spectra,
Aaron Dotter for helping with interpretation of the photometry, Alison
Doane at the Harvard College Observatory for helping us search the
plate stacks, Tony Misch at Lick Observatory for helping us date plates
taken by Paul Hodge, and John Grula at the Observatories of the
Carnegie Institution of Washington for locating several plates of
NGC~4449 taken at Mt.~Wilson which Francois Schweizer helped to
interpret. This research was supported in part by NSERC through a PGS
award, and by NASA through grants GO-6118 and GO-10286 from the Space
Telescope Science Institute, which is operated by the Association of
Universities for Research in Astronomy under contract NAS 5-26555.

\clearpage

\begin{deluxetable}{lcccc}
\centering
\tablecaption{Summary of {\it HST} ACS/WFC images}
\tablecolumns{5}
\tablewidth{0pt}
\tablehead{\colhead{Filter}                       &
           \colhead{$\lambda _{\rm cen}$}         &
           \colhead{$\lambda _{\rm FWHM}$}        &
           \colhead{Date}                         &
           \colhead{Exp Time}                    \\
           \colhead{}                             &
           \colhead{(\AA)}                        &
           \colhead{(\AA)}                        &
           \colhead{(UT)}                         &
           \colhead{(s)}                          }
\startdata
F435W & 4297 & 1038 & 10 Nov 2005 & 3660 \\
F502N & 5022 & 57   & 18 Nov 2005 & 1284 \\
F550M & 5580 & 547  & 18 Nov 2005 & 1200 \\
F555W & 5346 & 1193 & 10 Nov 2005 & 2460 \\
F658N & 6560 & 73   & 17 Nov 2005 & 1539 \\
F660N & 6602 & 35   & 18 Nov 2005 & 1860 \\
F814W & 8333 & 2511 & 10 Nov 2005 & 2060 \\
\enddata
\end{deluxetable}

\clearpage

\begin{figure}[hp]
\plotone{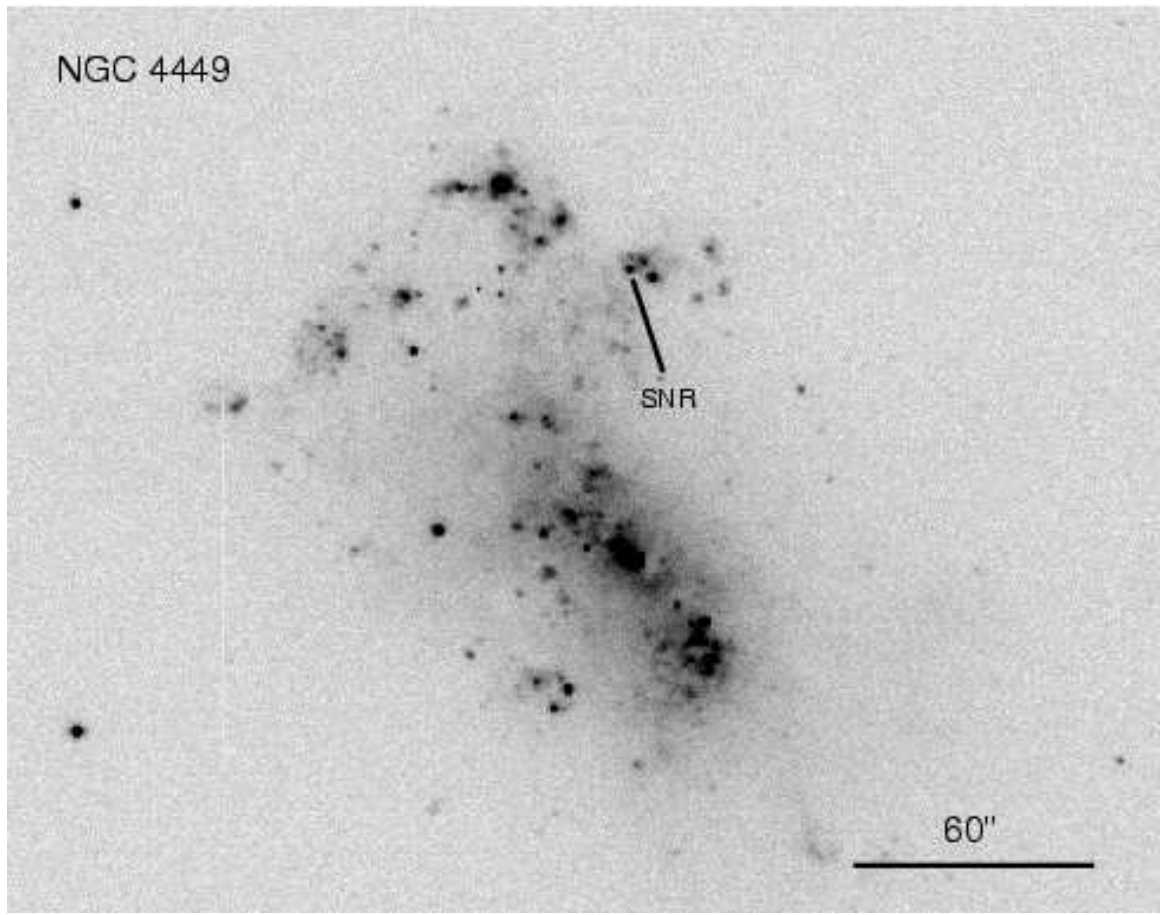}
\caption{An [\ion{O}{3}] 5007 \AA \ image of NGC~4449 obtained in 1996
  showing the location of the young, O-rich supernova remnant 1$'$
  north of the galaxy center. North is up and east is to the left.}
\end{figure}

\clearpage

\begin{figure}[hp]
\plotone{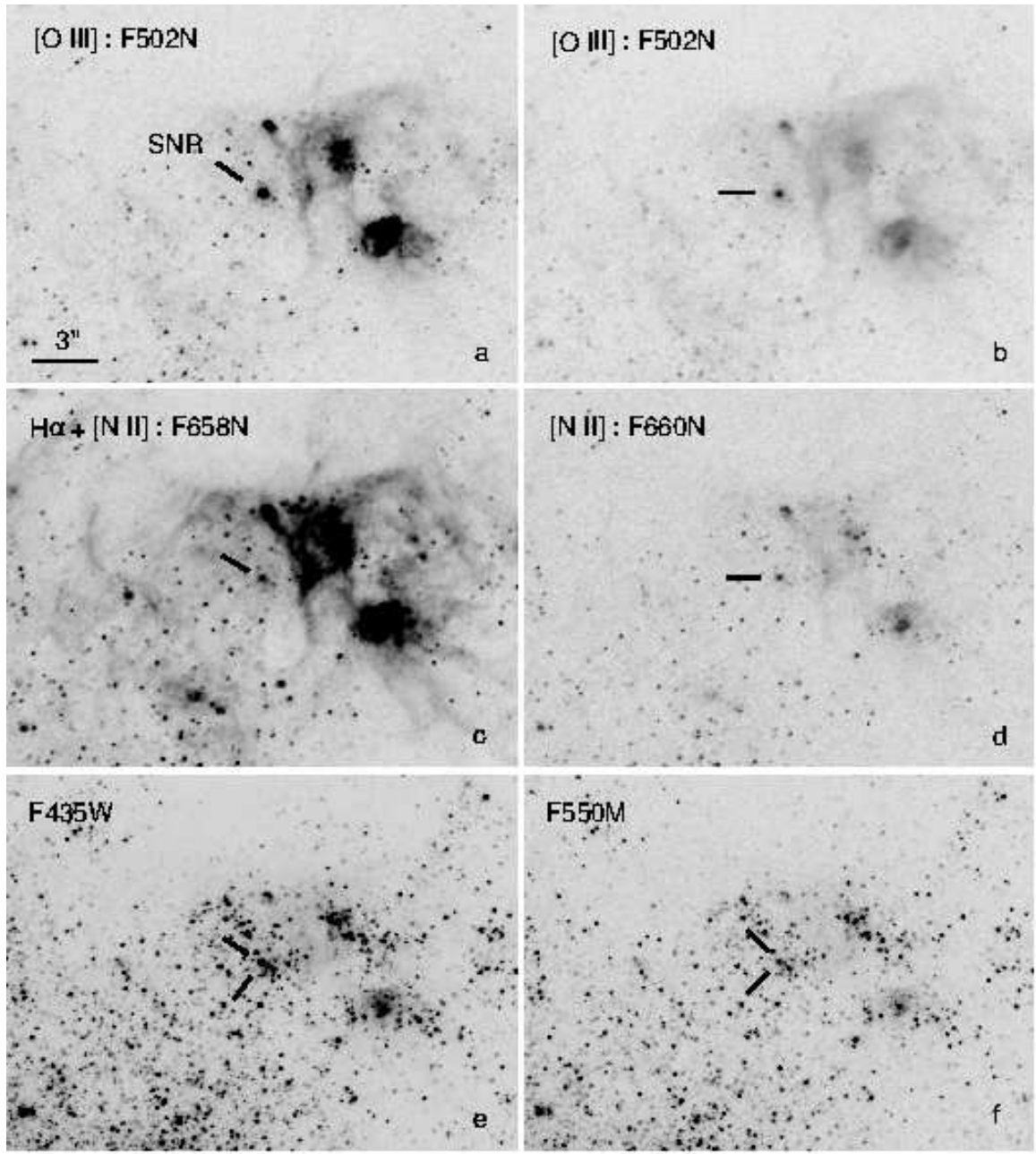}
\caption{{\sl HST} ACS/WFC images of the region surrounding
  SNR~4449-1 taken in 2005. (a) [\ion{O}{3}] $\lambda\lambda$4959,5007
  (F502N). (b) Same as in panel (a), but highlighting the brightest
  sources. (c) H$\alpha$ + [\ion{N}{2}] (F658N). (d) [\ion{N}{2}]
  $\lambda$6583 (F660N). (e) Blue (F435W). (f) Line-free continuum
  (F550M). North is up and east is to the left.}
\end{figure}

\clearpage

\begin{figure}[hp]
\plotone{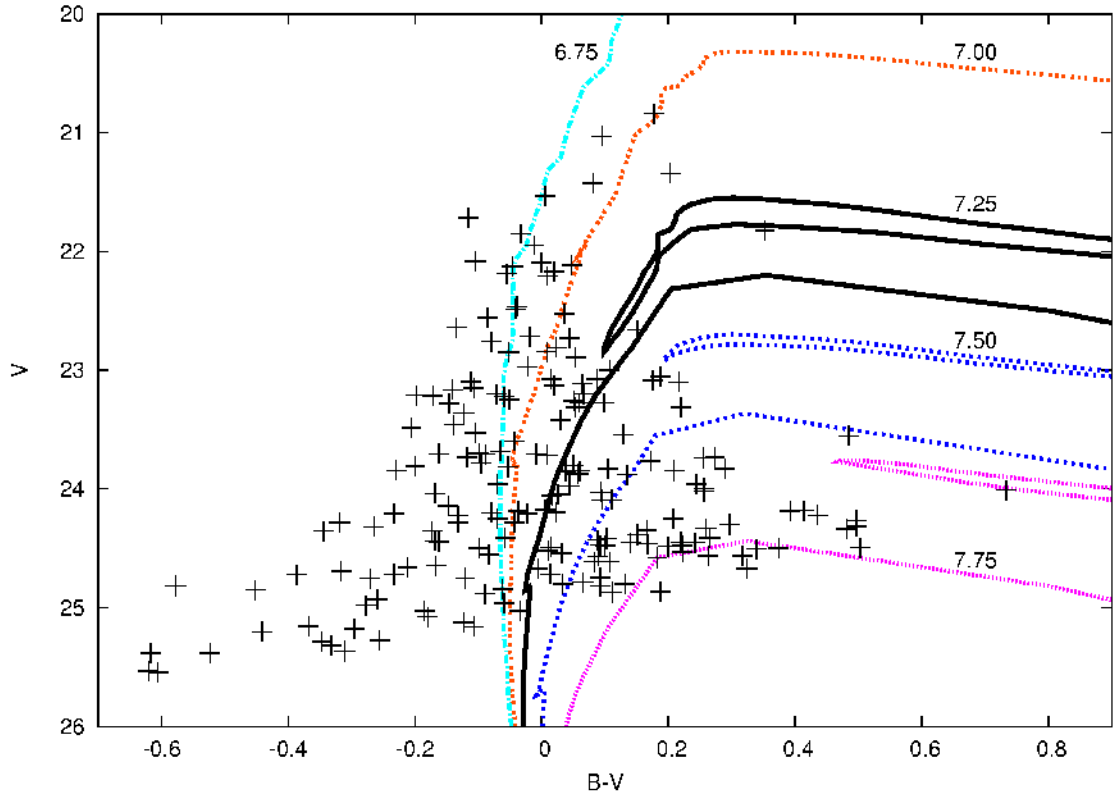}
\caption{A color-magnitude diagram of the stars in the
  immediate projected vicinity of SNR~4449-1.  Overlaid are
  isochrones of metallicity Z = 0.004 shifted to the
  distance, extinction and reddening of the remnant. The estimated
  minimum age of the stellar population is log (age yr$^{-1}$) = 7.25
  $\pm$0.50, consistent with the expected lifetime for stars with
  M$_{\rm ZAMS}$ = 15 -- 25 M$_{\odot}$.}
\end{figure}

\clearpage

\begin{figure}[hp]
\plotone{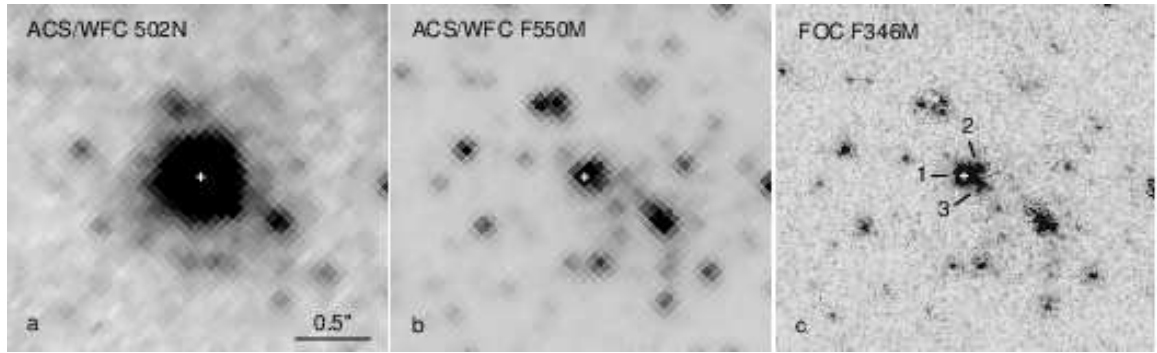}
\caption{{\it HST} images of the stellar environment of
  SNR~4449-1. (a) ACS/WFC [\ion{O}{3}] image with the ``+'' marking
  the center of the remnant's emission. (b) ACS/WFC line-free
  continuum image with the ``+'' marking the same location as the
  [\ion{O}{3}] image. (c) High resolution FOC F346M image.  Note the
  location of the remnant coincides with a tight grouping of luminous
  stars.  North is up and east is to the left.} 
\end{figure} 

\clearpage

\begin{figure}[hp]
\plotone{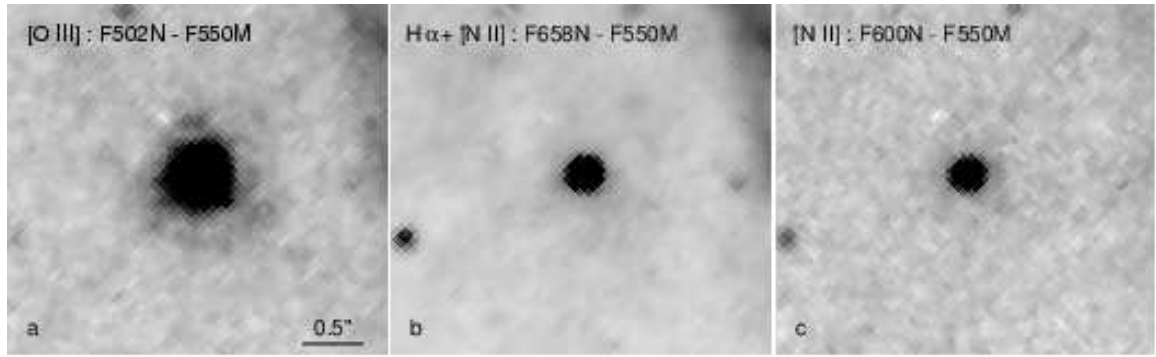}
\caption{Enlarged {\it HST} continuum-subtracted images of SNR~4449-1.
  (a) [\ion{O}{3}]-sensitive F502N image subtracted with a
  flux-calibrated and aligned F550M image. (b) H$\alpha$ +
  [\ion{N}{2}]-sensitive F658N image subtracted the same as in (a). (c)
  [\ion{N}{2}] 6583-sensitive F660N image subtracted the same as in
  (a).}
\end{figure}

\clearpage

\begin{figure}[hp]
\plotone{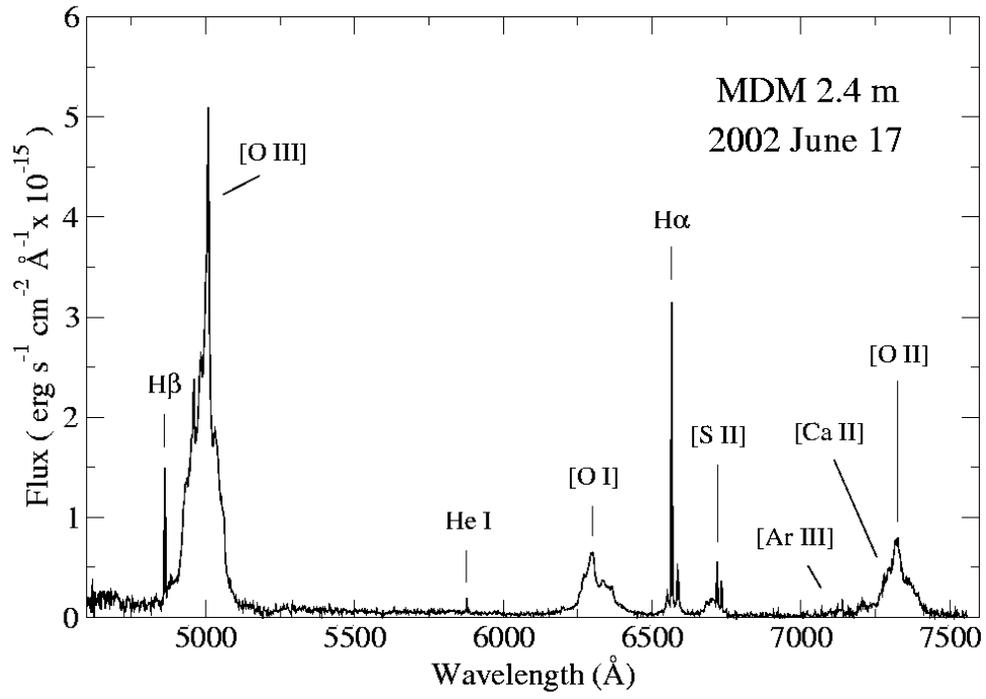}
\caption{Spectrum of SNR~4449-1 taken with the MDM 2.4 m telescope.}
\end{figure}

\clearpage

\begin{figure}[hp]
\plotone{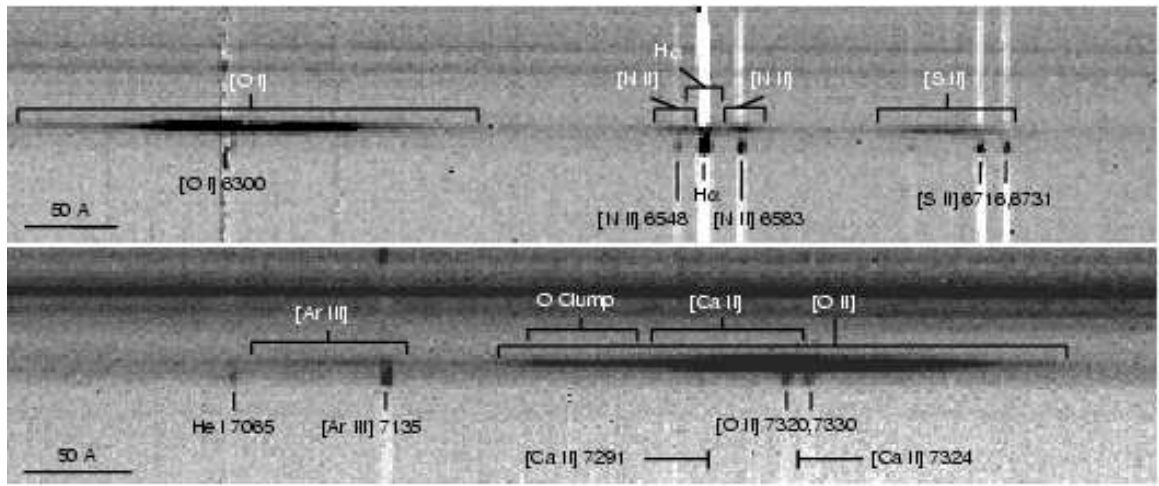}
\caption{2D background-subtracted spectra of SNR~4449-1 taken with the
  MDM 2.4 m telescope. The top spectrum
  was taken in 2002 and the bottom spectrum was taken in 2007. Emission lines in
  the rest frame of the galaxy are marked in black, while the white
  labels highlight broad features.}
\end{figure}

\clearpage

\begin{figure}[hp]
\plotone{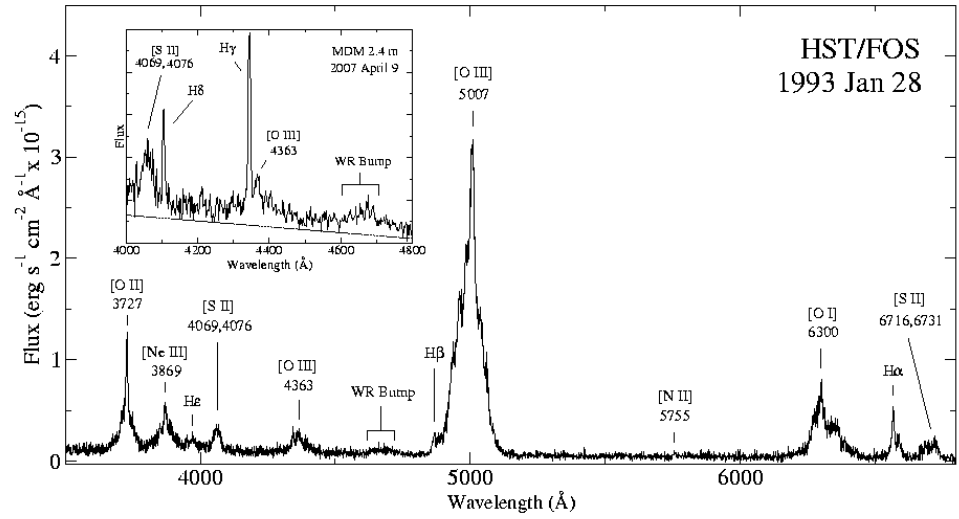}
\caption{{\it HST} FOS spectrum of SNR~4449-1.  The
  inset is a MDM spectrum that shares the
  WR-like broad emission at 4600 -- 4700 \AA.}
\end{figure}

\clearpage

\begin{figure}[hp]
\plotone{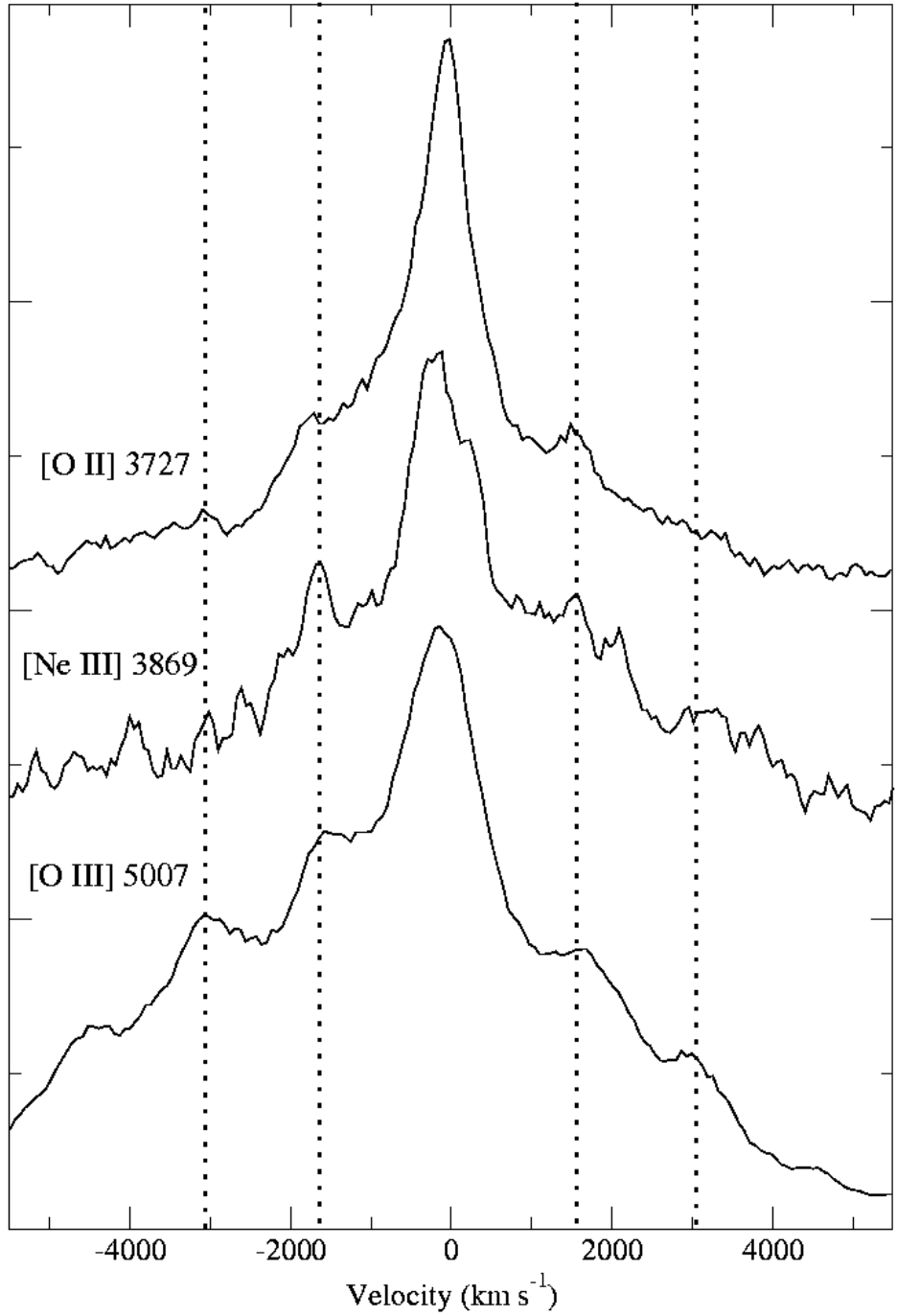}
\caption{Smoothed velocity line profiles of [\ion{O}{2}] 3727,
  [\ion{Ne}{3}] 3869 and [\ion{O}{3}] 4959,5007 from {\it HST} FOS
  spectra obtained in 1993.  Dotted lines are marked at $\pm$1600
  \kms. The [\ion{O}{3}] profile is composed of the 5007 line at 0
  \kms \ (marked ``A'') with associated blueshifted (A$_{\rm b}$) and
  redshifted  (A$_{\rm r}$) emission, and the 4959 line (marked ``B'')
  with associated blueshifted (B$_{\rm b}$) and redshifted (B$_{\rm
  r}$) emission. Velocities are in the rest frame of NGC 4449.}
\end{figure}

\clearpage

\begin{figure}[hp]
\plotone{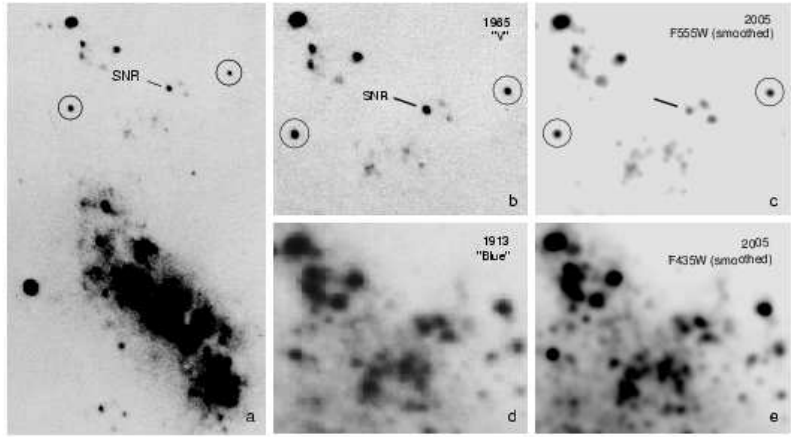}
\caption{{\it Left panel} (a): A 1965 plate of NGC~4449 taken with the
  Lick 3 m telescope using 103aD emulsion behind a GG 11 filter.
  Stellar sources are marked for brightness comparison.  (b) The same
  1965 image but enlarged around the remnant. (c) 2005 {\sl HST} F555W
  image smoothed to match the resolution of the 1965 image. (d) A 1913
  plate taken with the Mt.~Wilson 60 inch telescope using
  blue-sensitive emulsion. (e) 2005 {\sl HST} F435W image smoothed to
  match the 1913 image. North is up and east is to the left.}
\end{figure}

\end{document}